\documentclass[a4paper,fleqn,usenatbib]{mnras}

 \usepackage[normalem]{ulem}
 \usepackage[final]{pdfpages}
 
\usepackage[T1]{fontenc}
\usepackage{ae,aecompl}

\usepackage{graphicx}	
\usepackage{amsmath}	
\usepackage{amssymb}	
\usepackage{multirow}
\usepackage{multicol}
\usepackage{longtable}
\usepackage{supertabular}
\usepackage{natbib}
\usepackage{hyperref}

\title[ CH$_3$NCO in the ISM]{Methyl isocyanate CH$_3$NCO: An important missing organic in current astrochemical networks}

\author[Majumdar et al.]{
L. Majumdar$^{1,2}$\thanks{E-mail: liton.icsp@gmail.com}, J.-C. Loison$^{3,4}$, M. Ruaud$^{5}$, P. Gratier$^{1}$, V. Wakelam$^{1}$, A. Coutens$^{1}$ \\ 
$^{1}$ Laboratoire d'astrophysique de Bordeaux, Univ. Bordeaux, CNRS, B18N, allée Geoffroy Saint-Hilaire, 33615 Pessac, France\\
$^{2}$ Jet Propulsion Laboratory, California Institute of Technology, 4800 Oak Grove Drive, Pasadena, CA 91109, USA\\
$^{3}$ Univ. Bordeaux, ISM, UMR 5255, F-33400 Talence, France\\
$^{4}$ CNRS, ISM, UMR 5255, F-33400 Talence, France\\
$^{5}$ NASA Ames Research Center, Moffett Field, CA 94035, USA
}

\date{Accepted XXX. Received YYY; in original form ZZZ}

\pubyear{2015}

\begin{document}
\maketitle

\begin{abstract}
Methyl isocyanate (CH$_3$NCO) is one of the important complex organic molecules detected on the comet 67P/Churyumov-Gerasimenko by {\it Rosetta's} Philae lander. It was also 
detected in hot cores around high-mass protostars along with a recent detection in the solar-type protostar IRAS 16293-2422. We propose here a gas-grain chemical model to form CH$_3$NCO after 
reviewing various formation pathways with quantum chemical 
computations. We have used {\tt NAUTILUS} 3-phase gas-grain chemical model to compare observed abundances in the IRAS 16293-2422. Our chemical model clearly indicates the ice phase origin of CH$_3$NCO.
\end{abstract}




\section{Introduction}
Comets are considered to be the repository of the most pristine material from the origin of the solar system in the form of ice, dust, silicate and refractory organic material \citep{2011ARA&A..49..471M}. It is 
believed that some of the water and organic material found on Earth may have been delivered by comets \citep{2011Natur.478..218H}. In the Solar System, when a comet passes close to the Sun, it warms and begin to evaporate their surface and evolve gasses. This process, called outgassing, produces a coma of gas and dust that has been extensively observed \citep{2006IAUS..229..133C}. More than 20 organic molecules have been identified in the coma of 
comets via ground and space-based observations (\citet{2014A&A...566L...5B}; \citet {2004A&A...418.1141C}). The chemical composition of comets clearly indicates that these objects are populated with many organic compounds that are commonly detected 
in the ISM. \citet{2015SciA....1E0863B} have found a good correlation between the type of species detected in the coma of comets and those of warm molecular clouds. 

Recently, the spacecraft Rosetta has detected many complex organic molecules (COMs) (such as ethanol (CH$_3$CHO), formamide (NH$_2$CHO), methyl isocyanate (CH$_3$NCO), ethylamine (C$_2$H$_5$NH$_2$) and many more) on the 
material of the comet 67P/Churyumov-Gerasimenko by the COSAC experiment \citep{2015Sci...349b0689G} and even simplest amino acid glycine accompanied by methylamine and ethylamine in the coma  measured by the ROSINA \citep{2016SciA....2E0285A}. CH$_3$NCO is one of those organics that could play a role in the synthesis of amino acid chains called peptides \citep{Pascal2005}. It was first detected in Sgr B2 by \citet{2015ApJ...812L...5H} and later in Orion KL by \citet{2016A&A...587L...4C}. Recently, it was also detected in the solar-type protostar IRAS 16293-2422 by \citet{2017MNRAS.469.2230M} and  
\citet{2017MNRAS.469.2219L}. 

However, it is not well understood how CH$_3$NCO is formed in the ISM. Recently, \citet{2017MNRAS.469.2230M} has included the chemistry proposed by \citet{2015ApJ...812L...5H} and suggested that the production of CH$_3$NCO could occur mostly via the gas-phase chemistry after the evaporation of HNCO from grain surface. \citet{2017A&A...601A..49B} has considered a grain surface production of CH$_3$NCO in their model via the radical-addition reaction between CH$_3$ and OCN. Another study by \citet{2017MNRAS.469.2219L} claimed that CH$_3$NCO can be formed in the solid state by VUV irradiation of CH$_4$:HNCO mixtures through CH$_3$ and NCO recombinations. This motivated us to revisit the chemistry of CH$_3$NCO in the ISM.

This paper reports the first public gas-grain chemical network for CH$_3$NCO followed by astrochemical modelling of low mass protostar IRAS 16293. The chemistry is  presented in Section 2. The chemical model is described in Section 3 while the results are discussed in the last Section.

\section{Review of the interstellar chemistry of CH$_3$NCO}
Despite an intensive search, we did not find any reaction producing efficiently CH$_3$NCO in the gas phase. \citet{2015ApJ...812L...5H} proposed the following gas phase formation route:

\begin{eqnarray}
	\mathrm{HNCO + CH_3} &\longrightarrow& \mathrm{CH_3NCO + H}\\
	\mathrm{HNCO + CH_5^+} &\longrightarrow& \mathrm{CH_3NCOH^+ + H_2}\\
	\mathrm{CH_3NCOH^+ + e^-} &\longrightarrow& \mathrm{CH_3NCO + H}
\end{eqnarray}

But we have found that reaction 1 is endothermic by 77 kJ/mol and shows a Transition State (TS) located 
83 kJ/mol above the entrance level (see Table 4 in the online supplementary material) and thus cannot play any role in gas phase neither on grain surface at low temperature. Reaction 2 has been studied experimentally by \citet{doi:10.1021/j100457a004} and they observed proton transfer process which 
forms H$_2$NCO$^+$ and CH$_4$, instead of CH$_3$NCOH$^+$ and H$_2$. Reaction 3 has then no impact on the CH$_3$NCO formation.  An alternative could have been the CH$_3$NC + OH$\rightarrow$  H + CH$_3$NCO reaction, which is exothermic by 107 kJ/mol, 
and may not show any barrier. However, CH$_3$NC has a low abundance in molecular clouds (\citet{1988A&A...189L...1C}; \citet{2013A&A...557A.101G}). So the reaction 
CH$_3$NC + OH$\rightarrow$ H + CH$_3$NCO will involve very low fluxes and we can thus neglect this reaction.  

The detection of CH$_3$NCO being limited to warm sources (hot cores and hot corinos) suggests a formation of this molecule on grains rather than in the gas phase \citep{2016A&A...587L...4C}. We have found four potential grain surface reactions which may produce CH$_3$NCO efficiently. The first one is the s-CH$_3$ + s-OCN$\rightarrow$ s-CH$_3$NCO reaction (here `s' is to represent species on the surface). As it is a reaction between two radicals $^\bullet$CH$_3$ + $^\bullet$OCN, this reaction is barrier-less and leads to CH$_3$NCO which arises from the pairing up of electrons on the two reactants radicals.

The second one is the CH$_3$ + OCN$^{\mbox{-}}$$\rightarrow$ CH$_3$NCO + e$^{\mbox{-}}$ reaction which is exothermic by 30 kJ/mol at M06-2X/AVTZ level in the gas phase. The M06-2X highly nonlocal functional is 
developed by \citet{Zhao2008} and is well suited for structures and energetics of the transition states. OCN$^{\mbox{-}}$ is widely observed on interstellar ice \citep{2005A&A...441..249V} and CH$_3$ is supposed to be relatively abundant on ice and relatively mobile \citep{2017arXiv170106492W}, then this reaction may play a role in the CH$_3$NCO formation. However, the adsorption energy of OCN$^{\mbox{-}}$ on ice is large due to dipole-ion interaction, and even if free electron is also strongly bound to ice (\citet{2006JChPh.125g6101K}; \citet{2016PCCP...18.4652D}, this reaction may be endothermic on ice.  By comparison, the H + OCN$^{\mbox{-}}$$\rightarrow$  HNCO + e$^{\mbox{-}}$ reaction is more exothermic (by 107 kJ/mol in the gas phase) which should prevent OCN$^{\mbox{-}}$ detection on ice if this reaction was also exothermic on ice due to the importance of H reaction on ice. Then free electron should be less bounded on ice than OCN$^{\mbox{-}}$,  as a result CH$_3$ + OCN$^{\mbox{-}}$$\rightarrow$ CH$_3$NCO + e$^{\mbox{-}}$ reaction is likely endothermic on ice and is neglected here.

\begin{figure}
\centering
\includegraphics[width=0.5\textwidth]{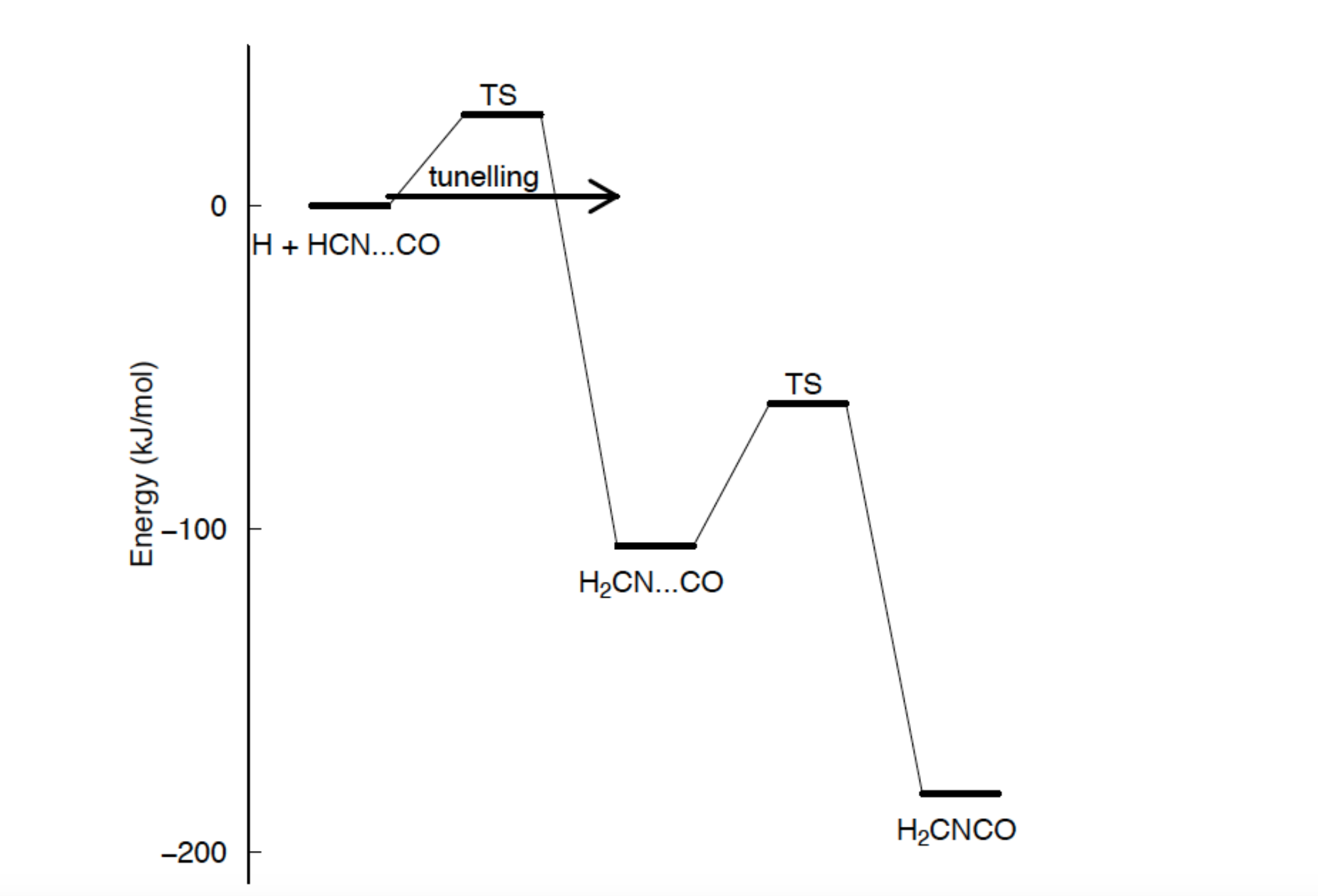}
\caption{Potential energy diagram for the H + HCN...CO reaction on the doublet surface
calculated at the M06-2X/aug-cc-pVTZ level including ZPE.}
\label{fig:diffDist}
\end{figure}

Another reaction is induced through the HCN...CO van der Waals formation on ice. As introduced in \citet{2015MNRAS.447.4004R}, the proximity of HCN and CO in the van der Waals complex favors the reaction between excited 
H$_2$CN{\bf *} (formed through s-HCN + s-H reaction) and CO (see Table 3 in the online supplementary material and Figure 1 for the energy profile diagram of the H + HCN...CO reaction). Then considering the 
large amount of CO and HCN, this reaction may be important. We then characterise the H + HCN and H$_2$CN + CO reactions to estimate the importance of these reactions to CH$_3$NCO formation on ice. The H + HCN$\rightarrow$ H$_2$CN shows a notable barrier (computed to be equal to 15.4 kJ/mol \citep{2013JChPh.139v4310J}, 28.3 kJ/mol at M06-2X/AVTZ level (this work, see Table 2 in the online supplementary material) and 36.4 kJ/mol at CCSD(T)/6-311++G level \citep{doi:10.1021/jp981542u}). Despite this barrier, this reaction is enhanced on ice due to tunneling \citep{2016MNRAS.459.3756R}, leading to the formation of excited H$_2$CN{\bf *}. As the amount of energy in H$_2$CN{\bf *} is equal to 105.3 kJ/mol (initially with a very narrow distribution), there is a competition between relaxation and reactivity for all the HCN linked to CO as the transition state for the H$_2$CN + CO$\rightarrow$ H$_2$CNCO is located 44.1 kJ/mol above the H$_2$CN + CO level, so 61.2 kJ/mol below the energy of the H$_2$CN{\bf *} formed through the H + HCN reaction. As the energy distribution of the H$_2$CN{\bf *} is initially very narrow, a notable part of the H$_2$CN{\bf *} will react with CO. It should be noted that this mechanism of reaction induced through van der Waals complex will also lead to HNCO formation through N...CO + H$\rightarrow$HNCO reaction, mechanism which should be very efficient in that case.

The fourth reaction which can produce CH$_3$NCO is the s-N + s-CH$_3$CO reaction. The first step, leading to s-CH$_3$C(N)O in a triplet state, is barrierless characteristic of a radical-radical reaction. The s-CH$_3$C(N)O can evolve toward s-CH$_3$ + s-OCN on the triplet surface, isomerize into s-CH$_3$NCO on the triplet surface or being converted into s-CH$_3$C(N)O in a singlet state (which can also isomerize into s-CH$_3$NCO (or s-CH$_3$OCN) on the singlet surface). Considering the exothermictiy of the various step (see Table 6 in the online supplementary material), the formation of s-CH$_3$NCO is without doubt the most favorable exit channel, either through adduct isomerization or through the recombination of s-CH$_3$ + s-OCN. Very minor CH$_3$OCN may also be produced, and some s-CH$_3$C(N)O may also be stabilized as the TS for dissociation or isomerization are notably above the CH$_3$C(N)O energy. s-CH$_3$C(N)O will react with s-H leading ultimately to CH$_3$C(O)NH$_2$ which are however not considered here. 

All the introduced and reviewed reactions discussed here are presented in Table 1.  

\begin{table*}
\scriptsize{
\label{list_reactions}
\caption{List of major gas-phase and grain surface reactions added to the model for CH$_3$NCO formation}
\begin{center}
\begin{tabular}{|clc|c|c|c|c|c|c|c|}
\hline
\hline
& Reaction  & & $\alpha$ & $\beta$ & $\gamma$ & Reference\\
\hline
1 & HNCO + CH$_3$ & $\rightarrow$  CH$_3$NCO + H  & $1.00\times 10^{-10}$ & 0 & $8.04\times 10^{3}$     & [1]\\

2 &  CH$_3$NCO + H$_3$$^+$  & $\rightarrow$    CH$_3$NCOH$^+$ +  H$_2$ & $1.00\times 10^{-9}$ & -0.5 & 0    & [2]\\

3 &  CH$_3$NCO + HCO$^+$  & $\rightarrow$    CH$_3$NCOH$^+$ +  CO & $1.09\times 10^{-9}$ & -0.5 & 0    & [2]\\

4 &  CH$_3$NCO + H$^+$  & $\rightarrow$    CH$_3$NCO$^+$ +  H & $1.00\times 10^{-9}$ & -0.5 & 0    & [2]\\

5 &  CH$_3$NCO + CO$^+$  & $\rightarrow$    CH$_3$NCO$^+$ +  CO & $1.00\times 10^{-9}$ & -0.5 & 0    & [2]\\

6 &  CH$_3$NCO + He$^+$  & $\rightarrow$    CH$_3$NCO$^+$ +  He & $1.00\times 10^{-9}$ & -0.5 & 0    & [2]\\

7 &  CH$_3$NCO$^+$ + e$^{\mbox{-}}$  & $\rightarrow$    CH$_3$ +  OCN & $1.50\times 10^{-7}$ & -0.5 & 0    & [3]\\

8 &  CH$_3$NCOH$^+$ + e$^{\mbox{-}}$  & $\rightarrow$    CH$_3$NCO +  H & $3.00\times 10^{-7}$ & -0.5 & 0    & [3]\\

9 &  CH$_3$NCO + CRP  & $\rightarrow$    CH$_3$ +  OCN & $4.00\times 10^{3}$ & 0 & 0    & [3]\\

10 &  CH$_3$NCO + Photon  & $\rightarrow$    CH$_3$ +  OCN & $5.00\times 10^{-10}$ & 0.0 & 0    & [3]\\

11 & HCN + s-CO & $\rightarrow$  s-HCN...CO & 1 & 0 & 0 & [4] \\

12 & s-HCN...CO + s-H & $\rightarrow$  s-H$_2$CNCO & 1 & 0 & $2.40\times 10^{3}$ & [5]   \\

13 & s-H$_2$CNCO + s-H  & $\rightarrow$  s-CH$_3$NCO & 1 & 0 & 0 & [6]\\

14 & s-CH$_3$ + s-HNCO & $\rightarrow$  s-CH$_3$NCO & 1 & 0 & $8.04\times 10^{3}$ & [1] \\

15 & s-CH$_3$ + s-OCN  & $\rightarrow$  s-CH$_3$NCO & 1 & 0 & 0 & [7]\\

15 & s-CH$_3$ + s-OCN$^{\mbox{-}}$  & $\rightarrow$  s-CH$_3$NCO + e$^{\mbox{-}}$ & 0 & 0 & 0 & [8]\\

16 & s-N + s-CH$_3$CO  & $\rightarrow$  s-CH$_3$NCO & 1 & 0 & 0 & [9]\\

\hline
\end{tabular}
\end{center}}
{\bf References}: [1] Current work and see Table 4 in the online supplementary material for detail calculation [2] \citet{1993JPCRD..22.1469A} [3] Considering the similar reactivity of HNCO   
[4] Following van der Waals formation on ice by \citet{2015MNRAS.447.4004R}  
[5] Current work and see Table 1 and 2 in the online supplementary material for detail calculation  
[6]  Current work and see Table 3 in the online supplementary material for detail calculation
[7] Following  \citet{2017A&A...601A..49B} 
[8] We neglect this reaction since we don't know exactly the value of the endothermicity.
 [9] Current work and see Table 6 in the online supplementary material for detail calculation \\
\end{table*}

\section{Astrochemical modelling}

\subsection{The NAUTILUS chemical model}
To investigate the chemistry of CH$_3$NCO in ISM, we have used the state-of-the-art chemical code {\tt NAUTILUS} described in \citet{2016MNRAS.459.3756R}. {\tt NAUTILUS} computes the chemical composition as a function of time 
in the gas-phase, and at the surface of dust grains. Here, surface chemistry is solved using the rate equation approximation and assuming a different 
chemical behaviour between the surface of the mantle and the bulk (i.e. a three-phase model). The equations and the
chemical processes included in the model are described in \citet{2016MNRAS.459.3756R}. {\tt NAUTILUS} includes physisorption of gas-phase species on grain surfaces, diffusion of species at the surface of the
grains resulting in chemical reactions and several desorption mechanisms. Desorption can be due to the temperature (thermal desorption), cosmic-ray heating
(cosmic-ray induced desorption, \citet{1993MNRAS.261...83H}), UV photon impact (photodesorption) and chemical \citep{2007A&A...467.1103G}. 

\subsection{Modification of the network}
Our gas phase chemistry is based on the public chemical network kida.uva.2014 \citep{2015ApJS..217...20W} with the updates on chemistry of HNCO (and their isomers) from the KIDA database\footnote{\url{http://kida.obs.u-bordeaux1.fr/}} and chemistry of CH$_3$NCO discussed in Section 2 and listed in Table 1 in this work. The surface network is based on the one of \citet{2007A&A...467.1103G} with
several additional processes from \citet{2015MNRAS.447.4004R} and Table 1 in this work. The entire network considered
here is available on the KIDA website.

\begin{figure}
\centering
\includegraphics[width=0.5\textwidth]{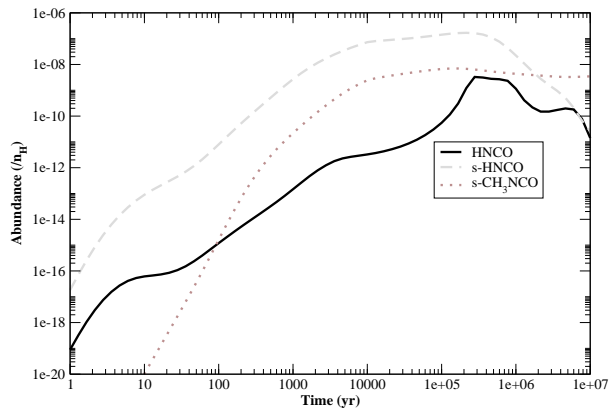}
\caption{  HNCO and CH$_3$NCO abundances (with respect to H) predicted by
our model for typical dense cloud conditions (see section 3.3) as a function of time in the gas-phase and at the surface
of the grains (represented by `s'). Here, abundance of gas phase CH$_3$NCO is negligible.}
\label{fig:diffDist}
\end{figure}

\subsection{Physical models}
To simulate the chemistry of CH$_3$NCO in the ISM, we have considered two different physical models which are representative of : (i) dense core and (ii) solar-type protostar (representative of IRAS 16293).
For dense core, {\tt NAUTILUS} is used with homogeneous conditions and integrated over 10$^7$ years. The initial elemental abundances are same as in \citet{2011A&A...530A..61H} with depleted value of fluorine abundance of 6.68$\times 10^{-9}$ from 
\citet{2005ApJ...628..260N} and we have used a standard C/O ratio of 0.7 (\citet{2015MNRAS.453L..48W}; \citet{2016MNRAS.458.1859M}). The model was run with constant dust and gas temperature of 10 K, a total proton density of $2\times10^{4}$ cm$^{-3}$, a cosmic-ray ionization rate of $1.3\times10^{-17}$ s$^{-1}$, and a visual extinction of 30.

For low mass protostar IRAS 16293, we have used the same physical structure as in \citet{2014MNRAS.445.2854W} and \citet{2016MNRAS.458.1859M} and that was computed using the radiation hydrodynamical (RHD) model from \citet{2000ApJ...531..350M}. 
Here the model initially starts from a dense molecular cloud core with a central density $n$(H$_2$) $\sim 3 \times 10^4$ cm$^{-3}$ and the core is extended up to  $r=4 \times 10^4$ AU with a total mass of 3.852 $M_{\odot}$. The prestellar core evolves to the protostellar core in $2.5 \times 10^5$ yr. When the protostar is formed, the model again follows the evolution for $9.3 \times 10^4$ yr, during which the protostar grows by mass accretion from the envelope.

\begin{figure*}
  \includegraphics[width=150mm]{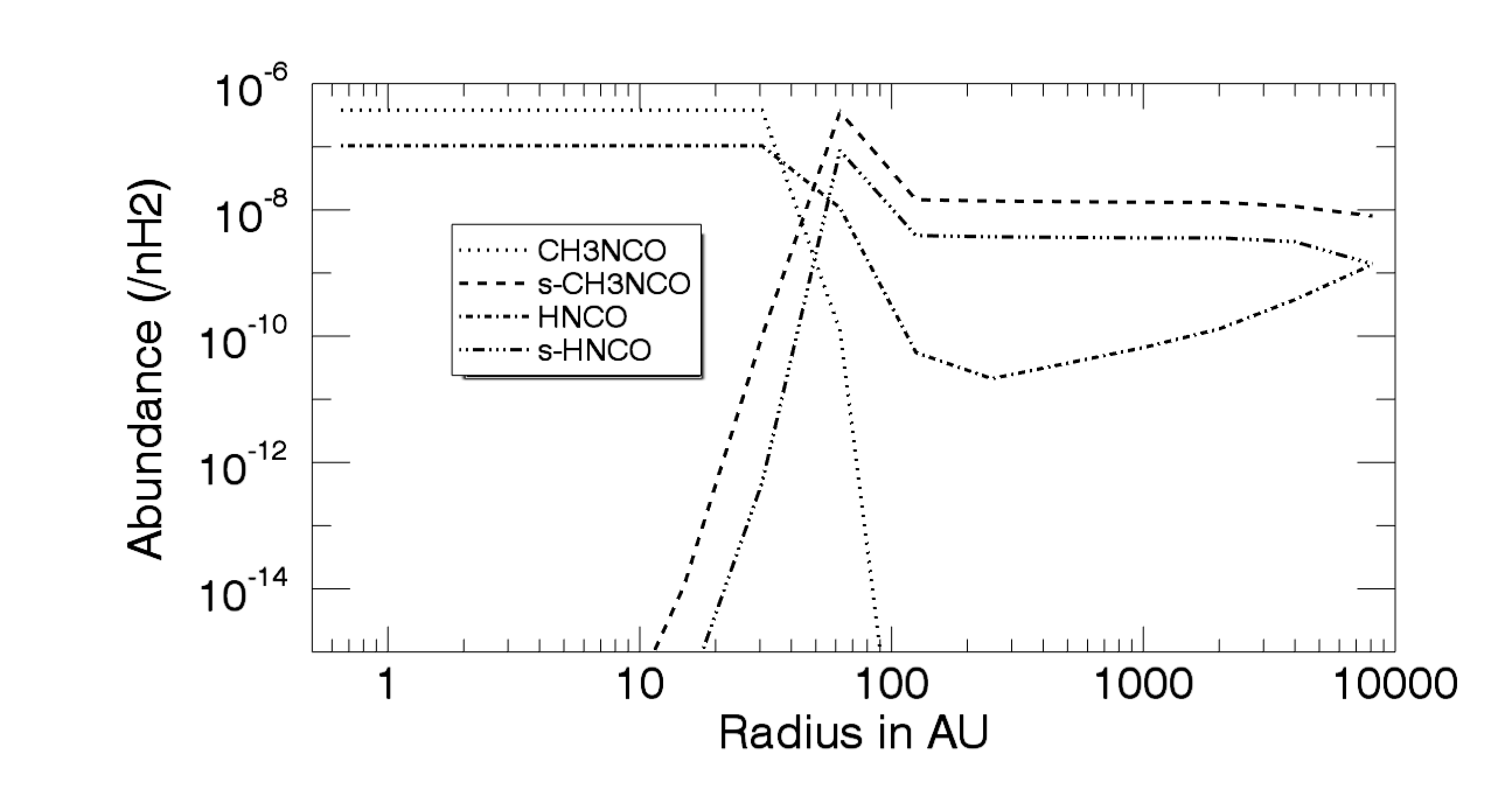}
  \caption{ Abundance with respect to H$_2$ predicted by our low mass protostar model for CH$_3$NCO and HNCO as a function of distance to the central star. s-CH$_3$NCO and s-HNCO represent CH$_3$NCO and HNCO on the grain surface.}
\end{figure*}

\section{Results and Discussions}

\subsection{Chemistry in dark cloud}
Figure 1 shows the time evolution of the abundances of HNCO and CH$_3$NCO under typical dark cloud conditions. Here, we have shown the chemical evolution of CH$_3$NCO together with HNCO since in the past HNCO was assumed to be the main precursor for CH$_3$NCO formation by \citet{2017MNRAS.469.2230M} and \citet{2015ApJ...812L...5H}. But this is not the case from our revised chemistry. \\

Recently, \citet{2016MNRAS.459.3756R} suggested a best fit chemical age for TMC-1 (CP) to be a few $10^5$ years.  At this time, 70\% of the observed species 
in TMC-1 (CP) are reproduced by the model within a factor of 10. At $2 \times10^5$ year, abundance of gas phase HNCO predicted by our model is $3\times10^{-10}$ with respect to n$_H$. This is in good agreement with 
with the observed abundance of $2\times10^{-10}$ in TMC-1 (CP) \citep{2013ChRv..113.8710A}. At this age, HNCO is formed mainly from the barrier-less surface reaction s-N + s-HCO and the gas phase dissociative recombination reaction of HNCOH$^+$.\\  

In our model, CH$_3$NCO is formed efficiently only in the surface via the reactions of s-H + s-H$_2$CNCO and s-N + s-CH$_3$CO since the main gas phase formation route CH$_3$ + HNCO has an large activation barrier of 8040 K. s-H$_2$CNCO is formed from the reaction between van der Waals complex s-HCN...CO and highly mobile s-H atoms. Here, s-HCN...CO is considered to be formed when gas phase HCN land on the proximity of s-CO on the grain surfaces \citep{2015MNRAS.447.4004R}. s-CH$_3$CO forms mainly via the reaction s-H + s-H$_2$CCO. The peak gas phase abundance of CH$_3$NCO from our model is of the order of $\sim10^{-32}$ with respect to n$_H$ whereas ice phase abundance is $7\times10^{-10}$ around few $10^5$ years. This shows that CH$_3$NCO is frozen in the ices around TMC-1 (CP) (also have lower rotational dipole moment compared to HNCO) and thus not free to rotate. Hence, CH$_3$NCO is not detectable via its millimeter wavelength rotational spectra around TMC-1 (CP). \\

\subsection{Chemistry in low mass protostar}
In order to validate our chemistry, we present the computed abundances of CH$_3$NCO, in the gas phase and at the surface of the grains in the protostellar envelope as a function of radius to the 
central protostar and compare with the observation of CH$_3$NCO in IRAS 16293. The abundance profiles of CH$_3$NCO can be discussed by considering two different regions. The 
first region is defined by radii larger than 200 AU where the temperature is below 50 K. In this region, most of the CH$_3$NCO in the grains is inherited from the cold core phase. The second region is 
defined by radii smaller than 100 AU where the temperature 
reaches above 100 K and the gas phase abundance of CH$_3$NCO increases sharply. The CH$_3$NCO abundance on grains has an inverse profile showing that at low temperature, the CH$_3$NCO molecules are formed on the 
grains and are thermally desorbed in the inner part of the envelope. 

\citet{2017MNRAS.469.2230M} has determined an upper limit of $1.4\times10^{-10}$ (with respect to molecular H$_2$) for CH$_3$NCO in the envelope of IRAS 16293 B at about 60 AU from the central protostar. Another 
study by \citet{2017MNRAS.469.2219L} has also determined an upper limit of $3.3\times10^{-10}$ (with respect to molecular H$_2$) for CH$_3$NCO in the same source IRAS 16293 B. Our model is in agreement with these upper limits 
at this radius ($2\times10^{-10}$). At 60 AU, the gas phase CH$_3$NCO/HNCO abundance ratio from our model is $2\times10^{-2}$ which is also reasonably close to the ratio $8\times10^{-2}$ measured by \citet{2017MNRAS.469.2230M} and \citet{2017MNRAS.469.2219L}. The ice phase CH$_3$NCO/HNCO ratio from our model in the outer part of the envelope (> 1000 AU) where possible comets should be formed is of the order of 3.7. This is very close to the CH$_3$NCO/HNCO$\sim$ 4.33 ratio initially measured by the Philae lander on the comet 67P/Churyumov-Gerasimenko \citep{2015Sci...349b0689G}. This observation is, however, now questioned (can only be considered as an upper limit) by the recent measurement from the Double Focusing Mass Spectrometer (DFMS) of the ROSINA experiment  \citep{2017MNRAS.469S.130A}.

\section{Conclusion and Perspectives} 
In this letter, we have provided new insights concerning the chemistry of CH$_3$NCO in the ISM. Our computation allowed us to confirm the hypothesis made by \citet{2016A&A...587L...4C} about its grain 
surface origin. Moreover, we tested the impact of these new kinetic data on the prediction of CH$_3$NCO abundance in the low mass protostar IRAS 16293-2422 and on the CH$_3$NCO/HNCO abundance ratio 
observed by the Philae lander on the comet 67P/Churymov-Gerasimenko. However, the study of the CH$_3$NCO/HNCO abundance ratio in high mass protostars with proper chemo-dynamical model is out of the 
scope of the current letter, but could be done in the near future by comparing with the observations in SgrB2 (N) (\citet{2015ApJ...812L...5H}; \citet{2017A&A...601A..49B}) and in Orion KL \citep{2016A&A...587L...4C}.

\section*{Acknowledgements}

LM, PG, VW and AC  thanks ERC starting grant (3DICE, grant agreement 336474) for funding during this 
work. LM also acknowledges partial support  from  the  NASA  postdoctoral  program. VW, PG and JCL acknowledge the French program Physique
et Chimie du Milieu Interstellair (PCMI) funded by the Conseil
National de la Recherche Scientifique (CNRS) and Centre National
d'Etudes Spatiales (CNES). A portion of this research was carried out at the Jet Propulsion Laboratory, California Institute of Technology, under a contract with the National Aeronautics and Space Administration.

\bibliographystyle{mnras}
\bibliography{CH3SH}

\clearpage

\newpage



\section*{Supplementary material}

**Appendix:**
A large part of the reactions treated in this study have unknown reaction rates (without experimental measurements and/or theoretical calculations to rely upon) in the temperature range of interest (T in the 10-50 K range) but also unknown exothermicity. To make a reasonable estimate of their reaction rates at low temperature, we performed quantum chemical calculations. For our calculations, we have used DFT method using the M06-2X functional with the cc-pVTZ basis set. This highly nonlocal functional developed by Zhao and Truhlar 2008 is well suited for structures and energetics of the transition states.

Table 1: Relative energies at the M06-2X/aug-cc-pVTZ level (in kJ/mol at 0 K including ZPE) with respect to the HCN + CO energy, geometries and frequencies (in cm$^{-1}$, unscaled) of the various stationary points. The absolute energies at the M06-2X/aug-cc-pVTZ level including ZPE in hartree are also given in column 1.

| Species (Energy, hartree) | Relative energies (kJ/mol) | Geometries (Position of the atoms in Cartesian coordinates) | | | Harmonic Frequencies (cm$^{-1}$) |
|---|---|---|---|---|---|
| | | x | y | z | |
| HCN (-93.407418) 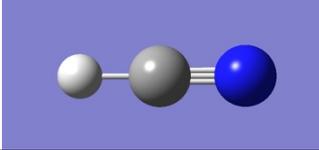 | 0 | C  0.000000<br>H  0.000000<br>N  0.000000 | 0.000000<br>0.000000<br>0.000000 | -0.494880<br>-1.560893<br>0.647167 | 789, 789, 2248, 3448 |
| CO (-113.315083) 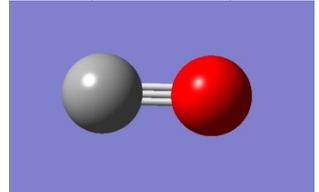 | | C  0.000000<br>O  0.000000 | 0.000000<br>0.000000 | -0.640800<br>0.480600 | 2260 |
| HCN…CO (-206.722951) 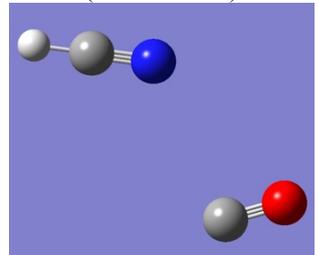 | -1.2 | C  -2.315172<br>H  -3.165133<br>N  -1.404659<br>O   2.154516<br>C   1.608775 | 0.203401<br>0.847465<br>-0.485993<br>-0.324565<br>0.655100 | 0.000202<br>0.000675<br>-0.000321<br>0.000274<br>-0.000305 | 22, 61, 67, 78, 791, 804, 2254, 2261, 3453 |

Table 2: Relative energies at the M06-2X/aug-cc-pVTZ level (in kJ/mol at 0 K including ZPE) with respect to the H + HCN energy, geometries and frequencies (in cm$^{-1}$, unscaled) of the various stationary points. The absolute energies at the M06-2X/aug-cc-pVTZ level including ZPE in hartree are also given in column 1.

| Species (Energy, hartree) | Relative energies (kJ/mol) | Geometries (Position of the atoms in Cartesian coordinates) | | | Harmonic Frequencies (cm$^{-1}$) |
|---|---|---|---|---|---|
| | | x | y | z | |
| H (-0.4982065) | 0 | | | | |
| HCN (-93.407418) 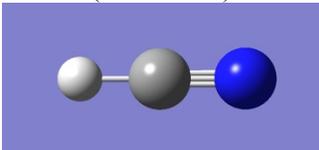 | | C 0.000000<br>H 0.000000<br>N 0.000000 | 0.000000<br>0.000000<br>0.000000 | -0.494880<br>-1.560893<br>0.647167 | 789, 789, 2248, 3448 |
| TS (-93.894842) 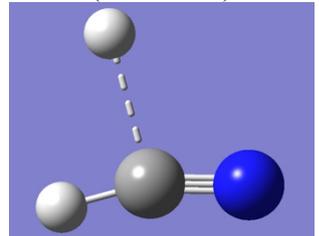 | 28.3 | N 0.097054<br>C 0.097054<br>H 0.394533<br>H -1.656237 | -0.717467<br>0.436285<br>1.463794<br>0.940767 | 0.000000<br>0.000000<br>0.000000<br>0.000000 | 505, 779, 921, 2134, 3388, 889i* |
| H$_2$CN (-93.945717) 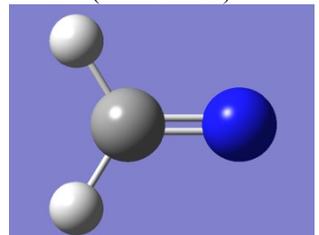 | -105.3 | C 0.000000<br>N 0.000000<br>H 0.000000<br>H 0.000000 | 0.000000<br>0.000000<br>0.937025<br>-0.937025 | -0.501313<br>0.735111<br>-1.068947<br>-1.068947 | 948, 1012, 1384, 1749, 3005, 3072 |

*: imaginary frequency

Table 3: Relative energies at the M06-2X/aug-cc-pVTZ level (in kJ/mol at 0 K including ZPE) with respect to the H$_2$CN + CO energy, geometries and frequencies (in cm$^{-1}$, unscaled) of the various stationary points. The absolute energies at the M06-2X/aug-cc-pVTZ level including ZPE in hartree are also given in column 1.

| Species (Energy, hartree) | Relative energies (kJ/mol) | Geometries (Position of the atoms in Cartesian coordinates) x y z | Harmonic Frequencies (cm$^{-1}$) |
|---|---|---|---|
| H$_2$CN (-93.945717) 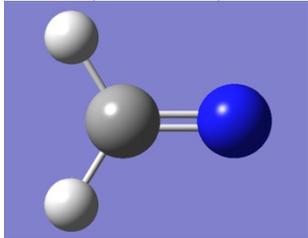 | 0 | C  0.000000   0.000000  -0.501313<br>N  0.000000   0.000000   0.735111<br>H  0.000000   0.937025  -1.068947<br>H  0.000000  -0.937025  -1.068947 | 948, 1012, 1384, 1749, 3005, 3072 |
| CO (-113.315083) 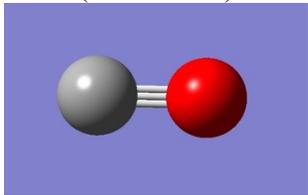 | | C  0.000000   0.000000  -0.640800<br>O  0.000000   0.000000   0.480600 | 2260 |
| TS (-207.244021) 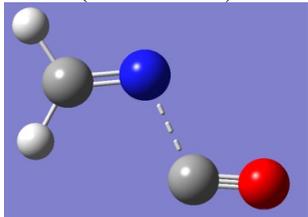 | 44.1 | C  1.719507   0.198906   0.000021<br>H  2.715274  -0.248015   0.000178<br>H  1.639585   1.285638  -0.000032<br>N  0.731882  -0.567158  -0.000051<br>O -1.840088  -0.111857   0.000040<br>C -0.845729   0.438984  -0.000040 | 99, 294, 442, 5459, 1062, 1082, 1415, 1722, 461i* |
| CH$_2$NCO (-207.290007) 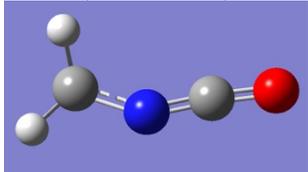 | -76.7 | C -1.778814   0.139360   0.000083<br>H -1.993176   1.195587  -0.000180<br>H -2.555836  -0.603926   0.000319<br>N -0.496439  -0.287279  -0.000136<br>C  0.672696  -0.037344  -0.000042<br>O  1.832599   0.100900   0.000071 | 134, 261, 417, 544, 629, 974, 1133, 1490, 1557, 2410, 3175, 3309 |

*: imaginary frequency

Table 4: Relative energies at the M06-2X/aug-cc-pVTZ level (in kJ/mol at 0 K including ZPE) with respect to the $CH_3$ + HNCO energy, geometries and frequencies (in cm$^{-1}$, unscaled) of the various stationary points. The absolute energies at the M06-2X/aug-cc-pVTZ level including ZPE in hartree are also given in column 1.

| Species (Energy, hartree) | Relative energies (kJ/mol) | Geometries (Position of the atoms in Cartesian coordinates) | | | Harmonic Frequencies (cm$^{-1}$) |
|---|---|---|---|---|---|
| | | x | y | z | |
| $CH_3$ (-39.795501) 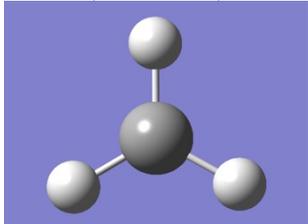 | 0 | C  0.000000<br>H  0.000000<br>H  0.932296<br>H -0.932296 | 0.000000<br>1.076523<br>-0.538261<br>-0.538261 | 0.000000<br>0.000000<br>0.000000<br>0.000000 | 524, 1416, 1416, 3133, 3310, 3310 |
| HNCO (-168.665769) 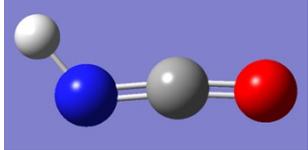 | | N  1.154466<br>C -0.045063<br>O -1.204109<br>H  1.821983 | -0.122413<br>0.016768<br>0.015868<br>0.629339 | 0.000026<br>-0.000079<br>0.000031<br>0.000041 | 567, 660, 787, 1371, 2369, 3698 |
| TS (-208.429659) 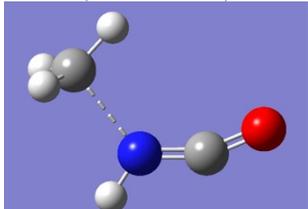 | 83.0 | C -0.920547<br>O -1.786482<br>N  0.248291<br>H  0.499206<br>C  1.895977<br>H  1.460329<br>H  2.432240<br>H  2.309463 | 0.289264<br>-0.500823<br>0.696949<br>1.668765<br>-0.433809<br>-1.413255<br>-0.036324<br>-0.223972 | 0.000721<br>-0.000027<br>-0.001215<br>0.005455<br>0.000426<br>-0.130380<br>-0.848907<br>0.975670 | 101, 159, 252, 551, 591, 669, 674, 969, 1003, 1241, 1432, 1441, 2197, 3109, 3262, 3277, 3704, 747i* |

*: imaginary frequency

Table 5: Relative energies at the M06-2X/aug-cc-pVTZ level (in kJ/mol at 0 K including ZPE) with respect to the N + CH$_3$CO energy, geometries and frequencies (in cm$^{-1}$, unscaled) of the various stationary points. The absolute energies at the M06-2X/aug-cc-pVTZ level including ZPE in hartree are also given in column 1.

| Species (Energy, hartree) | Relative energies (kJ/mol) | Geometries (Position of the atoms in Cartesian coordinates) | | | Harmonic Frequencies (cm$^{-1}$) |
|---|---|---|---|---|---|
| | | x | y | z | |
| N($^4$S) (-54.58746) | | | | | |
| CH$_3$CO($^2$A') (-153.127903) 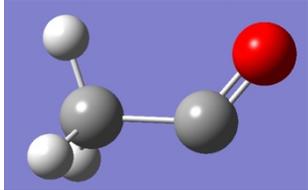 | 0 | C -0.388178<br>O 0.332527<br>C -0.032493<br>H -0.490106<br>H -0.470486<br>H 1.049276 | 0.005985<br>-0.003169<br>0.000250<br>-0.873754<br>0.882909<br>-0.012223 | 0.047915<br>0.973347<br>-1.417739<br>-1.877220<br>-1.879688<br>-1.552598 | 62, 469, 867, 958, 1051, 1355, 1460, 1460, 1989, 3050, 3149, 3151 |
| CH$_3$C(N)O($^3$A) (-207.821775) 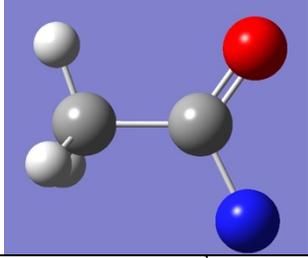 | -279 | O -0.867949<br>C 1.345357<br>H 1.717670<br>H 1.717617<br>H 1.696968<br>C -0.156245<br>N -0.760476 | -1.010897<br>-0.042331<br>0.484667<br>0.482425<br>-1.069688<br>-0.028182<br>1.230407 | -0.000001<br>-0.000012<br>-0.878557<br>0.879906<br>-0.001305<br>0.000018<br>-0.000010 | 85, 380, 518, 522, 844, 942, 1023, 1234, 1389, 1467, 1470, 1692, 3063, 3132, 3176 |
| CH$_3$C(N)O($^1$A) (-207.815884) 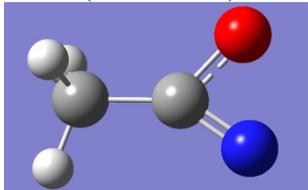 | -264 | O -0.961115<br>C 1.473599<br>H 1.777168<br>H 1.934804<br>H 1.777450<br>C 0.013128<br>N -0.960123 | -0.798213<br>-0.015759<br>-0.577365<br>0.967298<br>-0.575912<br>0.099415<br>0.867106 | -0.000009<br>-0.000021<br>-0.882612<br>-0.000847<br>0.883383<br>0.000068<br>-0.000020 | 141, 361, 434, 544, 913, 981, 1056, 1255, 1404, 1466, 1478, 1944, 3076, 3154, 3173 |
| TS($^3$A) (→ CH$_3$ + NCO) (-207.774031) 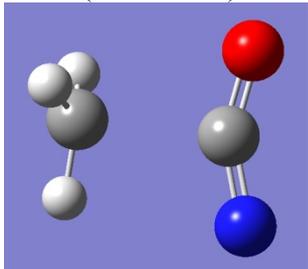 | -154 | O -0.651718<br>C 1.572500<br>H 1.600004<br>H 1.804223<br>H 1.844574<br>C -0.503234<br>N -0.921522 | -1.175499<br>0.110376<br>1.188300<br>-0.305225<br>-0.474905<br>0.013874<br>1.178618 | 0.000121<br>0.000428<br>-0.099523<br>0.969591<br>-0.865333<br>-0.001984<br>0.000519 | 127, 265, 443, 509, 653, 673, 992, 1252, 1412, 1432, 1814, 3105, 3263, 3293, 503i* |
| TS($^1$A) (CH$_3$OCN → CH$_3$NCO) (-207.783064) 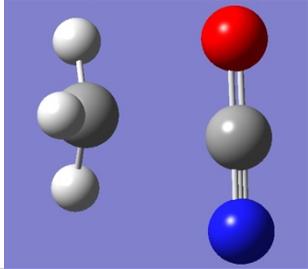 | -178 | O -0.285154<br>C -1.114689<br>H -0.711470<br>H -1.174977<br>H -1.875839<br>C 0.555497<br>N 1.342667 | 1.199652<br>-0.748877<br>-1.598000<br>-0.822782<br>-0.171890<br>0.280798<br>-0.599438 | -0.001644<br>0.001882<br>-0.526160<br>1.076861<br>-0.495240<br>-0.009050<br>0.000101 | 135, 454, 535, 606, 848, 912, 1218, 1297, 1432, 1468, 2092, 3151, 3318, 3328, 860i* |
| CH$_3$NCO($^1$A) (-207.933895) | -574 | | | | 56, 151, 622, 659, 895, 1132, 1160, 1459, 1498, |

| | | | | 1517, 1552, 2415, 3061, 3129, 3154 |
|---|---|---|---|---|
| CH$_3$OCN($^1$A) (-207.891329) 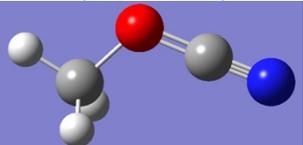 | -462 | C  1.480716   0.349286   0.000000<br>H -2.401061  -0.222819  -0.000183<br>H -1.408142   0.963121   0.894860<br>H -1.407932   0.963363  -0.894678<br>C  0.767246  -0.133253   0.000001<br>N  1.835008   0.294098   0.000000<br>O -0.418388  -0.632318   0.000000 | | 154, 235, 551, 644, 938, 1170, 1178, 1252, 1479, 1503, 1503, 2420, 3081, 3166, 3203 |
| CH$_3$ (-39.795501) 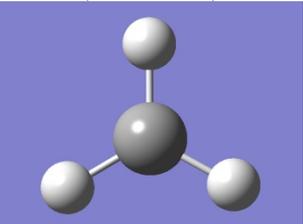 | -188 | C  0.000000   0.000000   0.000000<br>H  0.000000   1.076523   0.000000<br>H  0.932296  -0.538261   0.000000<br>H -0.932296  -0.538261   0.000000 | | 524, 1416, 1416, 3133, 3310, 3310 |
| NCO (-167.991408) 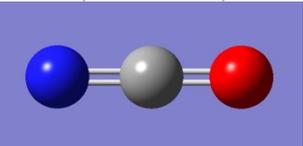 | | N  0.000000   0.000000  -1.259704<br>C  0.000000   0.000000  -0.037349<br>O  0.000000   0.000000   1.130252 | | 545, 619, 1330, 2053 |

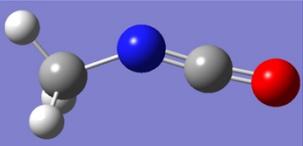

*: imaginary frequency



\end{document}